# Analyzing the Relationship among Cryptocurrencies using Complex Networks

Stergios Intzes[1], Georgios D. Papadopoulos[2], Lykourgos Magafas[3]

1  Department of Management Science and Technology, Democritus University, Kavala, Greece

2  Department of Physics, Faculty of Sciences, Democritus University, Kavala, Greece

3  Department of Informatics, Democritus University, Kavala, Greece

e-mail: sintzes@mst.duth.gr  geopapado@physics.duth.gr  magafas@cs.duth.gr

*Abstract*

Our analysis focuses on the stock cryptocurrency market, by studying a group of nineteen cryptocurrencies where their capitalization is about 99% of the total market. Specifically, it is examined this group of cryptocurrencies for the period from 2017 up to 2024, taking into account the effect of COVID-19 pandemic. As far as we know, this is the first time that this kind of study has been published, where it takes place by creating various networks based on stock cryptocurrency correlation, in order to be possible to visualize and access these relationships using methods from Complex Networks. The evaluation results show that, there are three different communities within the crypto market and although COVID-19 pandemic and ongoing geopolitical changes, there is a notable trend towards an increase in the number of cryptocurrencies. Additionally, it is applied discriminant analysis to identify the differences among various cryptocurrencies based on their features.

**1.Introduction**

Cryptocurrencies represent a significant innovation in the field of digital finance, leveraging blockchain technology to offer secure, and transparent financial transactions[1]. On October 31, 2008 a computer programmer, with the pseudonym Satoshi Nakamoto, presented the creation of bitcoin as a "new electronic cash system" that would be based on the blockchain method [2], he could hardly have imagined the exponential growth of its market capitalization in next years. We aim to utilize the principles of complex systems in order to discern the characteristics of cryptocurrencies and the interconnections between them, using the daily fluctuations [3, 4].

More specifically we seek to build network models that demonstrate relationships between cryptos. As raw data we use share's daily closing prices from the each crypto and compute the related interaction between them. Our analysis is focused on a group of nineteen cryptocurrencies, which collectively account for approximately 99% of the total market capitalization from 2017 to the present day (20 May 2024) .Complex Networks will be used in order to draw conclusions regarding the topology and in finding the most central shares [5]. For this purpose, we used the open-source network analysis software "Gephi" [6] and we develop a series of questions.

Q1: Are there relationships among cryptocurrencies?

Q2: What impact has COVID-19 on the cryptocurrency ecosystem?

Q3: Verifying whether Bitcoin is associated with specific communities

Additionally, we perform discriminant analysis [7] in order to analyzing crypto's data, facilitate dimensionality reduction and finally feature extraction. By developing discriminant functions and deriving cutoff scores, discriminant analysis helps us in accurately predicting and classifying samples of cryptos into distinct groups [8].

**2. Theoretical and Methodological Issues**

Definition: A network (or graph) is a collection of nodes (also called vertices) and edges (also called links) that connect pairs of nodes. Formally, a network G is defined as an ordered pair G= (V, E) where:

- V is a set of nodes (vertices).
- E is a set of edges (links) that connect pairs of nodes.

The total number of nodes is denoted by n (n = |V|). Set E contains ordered or unordered pairs of elements on $V(i,j)$, $i, j \in V$, are called arcs or links respectively. It is simple to create a visual representation using a graph to illustrate important connections. In this type of diagram, a node is represented as a point, circle, or rectangle that includes the node's label (if nodes are unnamed, integers can be used instead). Arcs are depicted as arrows pointing in a specific direction, while links are shown as straight lines connecting the nodes involved.

1. Average Degree: The average degree ⟨k⟩ of the network is calculated as the mean degree over all nodes in the network. For an undirected network, the average degree is given by: $Average\ Degree = \frac{Total\ Edges}{Total\ Nodes}$
2. Graph Density: The density of a graph is calculated by dividing the total number of existing arcs by the maximum number of potential arcs within the graph. A complete graph contains all possible pairs of arcs, resulting in a density of 1 [9, 10]. Conversely, a completely disconnected graph with no arcs present has a density of 0. A graph is considered dense when its density approaches 1, while it is classified as sparse when the density is lower. In an undirected graph with N nodes, the maximum number of links that can exist is:

$$\frac{N(N-1)}{2}$$

Hence, if this graph has L links, then its density S, is calculated as: $S = \frac{L}{N(N-1)}$
3. Network Diameter: The diameter of a network is a key metric in the analysis of complex networks. It provides a measure of the network's overall size in terms of the maximum distance between any pair of nodes [9]. This metric is crucial for understanding the efficiency of information or resource transfer across the network. The diameter of a network is defined as the greatest shortest path length between any pair of nodes in the network. Formally, for a network G= (V, E) the diameter D is given by: $D = max_{u,v \in V}\ d(u,v)$, where $d(u,v)$ represents the shortest path length between nodes u and v.
4. Average Path Length: Average path length [11] $d$ of the network is the average directed distance between any pair of nodes, indicating the efficiency of information or mass transport on a network.
5. Communities: Complex networks have community structure [12] if the nodes of the networks can be grouped into sets of nodes where each set of nodes is densely connected, at least, internally.

## 3. Methodology and Analysis

*3.1. Data Collection and Networks*

The plan was to collect all daily closing prices for nineteen cryptos where their capitalization is about 99% of the total market. Our research focuses on the years 2017 to 2024, with a specific emphasis on the impact of the SARS-CoV-2 pandemic, aiming to analyze differences in behaviors and the influence of the pandemic on the cryptocurrency market[13]. Finally, using Gephi software, these data will produce networks which will be visualized and analyzed with respect to topology of networks[14]. To identify the most reliable data sources, we opted to utilize the website focused on economic matters (https://www.investing.com/).

Also, Pearson correlation [15] is a measure of the linear relationship between two variables. It is used in statistics to quantify the strength and direction of the association between two continuous variables.

Additionally, we calculated the daily logarithmic changes [12] of each stock, denoted as I, for every trading day of the year using the following equation:

$$r_i(\tau) = lnP_i(\tau) - lnP_i(\tau - 1)$$

where $P_i(\tau)$ is the closing price of stock $i$ on day $\tau$. As a result, the Pearson Correlation is calculated for every pair of two cryptocurrencies, leading to the creation of a 19x19 matrix. As the correlation value of two stocks nears 1, they tend to move in the same direction. Conversely, when the correlation coefficient approaches -1, the stocks move in opposite directions. A correlation coefficient of zero indicates that there is no predictable relationship between the movements of the two stocks.

Consequently, we developed the network of all cryptocurrencies for the period from 2017 up to 2024 and it was calculated the number of communities, the average degree, as well as the total number of nodes and edges (see Figure 1). It is obvious that during the period from 2021 to 2022, coinciding with the COVID-19 pandemic, there was an increase in the number of communities. It was found three (Figure 1C) different communities, and it can be asserted that there are three separate clusters of changes, which vary according to the strength of the correlation.

Additionally, the average degree has been on the rise, paralleling the growth in the number of nodes and edges. Despite the lockdown and the ongoing conflict in Ukraine, the number of cryptos continues to rise, which indicates that investors have trust in the cryptocurrency market. It is a matter of concern, as we can see that the indices of international stock markets tend to decline during times of war or pandemic. Also, it can be observed that the distribution of edges appears to follow an exponential distribution.

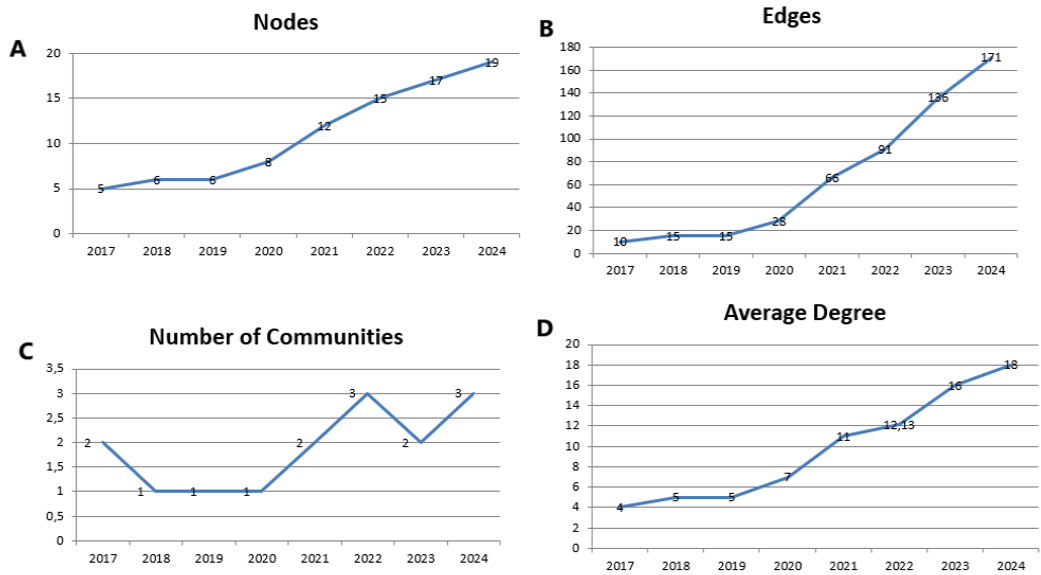

*Figure 1* Time series of topological metrics of all networks from 2017 to 2024. (A) Number of Nodes (B) Number of Edges (C) Number of Communities (D) Average Degree

Moreover, we calculate all metrics of networks, closeness centrality, betweenness centrality, eigenvector centrality (Table 1) which present an increasing trend over time. It is noticed that, high betweenness centrality might be influencers or connectors between different communities, facilitating the spread of information. Moreover, an increase in eigenvector centrality may suggest an expansion of the network, as it signifies a growing number of users or entities over time.

|  | 2017 | 2018 | 2019 | 2020 | 2021 | 2022 | 2023 | 2024 |
|---|---|---|---|---|---|---|---|---|
| Nodes | 5 | 6 | 6 | 8 | 12 | 15 | 17 | 19 |
| Edges | 10 | 15 | 15 | 28 | 66 | 91 | 136 | 171 |
| eigengAverage Degree | 4 | 5 | 5 | 7 | 11 | 12,133 | 16 | 18 |
| Network Diameter | 1 | 1 | 1 | 1 | 1 | 1 | 1 | 1 |
| Graph Density | 1 | 1 | 1 | 1 | 1 | 0,867 | 1 | 1 |
| Avg. Path Length | 1 | 1 | 1 | 1 | 1 | 1 | 1 | 1 |
| Number of Communities | 2 | 1 | 1 | 1 | 2 | 3 | 2 | 3 |
| Closeness Centrality |  | 5 | 6 | 6 | 8 | 12 | 14 | 17 | 19 |
| Betweenness Centrality | 5 | 6 | 6 | 8 | 12 | 15 | 17 | 19 |
| Eigenvector Centrality | 5 | 6 | 6 | 8 | 12 | 14 | 17 | 19 |

*Table 1* Metrics from all crypto networks from 2017 to 2024.

We applied the Vincent D Blondel algorithm [16] to illustrate the communities that were formed. It is evident that Bitcoin is part of the second community, along with other cryptocurrencies that share strong connections. For instance, Bitcoin and Ethereum exhibit a Pearson correlation of 0.734, while for the case of BNB and Wrapped Bitcoin is 0.773. The other communities consist from cryptos with lower Pearson correlation (figure 2).

The implementation of the algorithm enables us to observe the progress within various communities, which provides valuable information of nature of interactions and behaviors within the cryptocurrency ecosystems. This indicates that investors exhibit three different clusters of recognition regarding cryptocurrencies, perhaps by the level of their fame. It is important to note that, crypto assets are decentralized, as they are not issued or governed by any central authority or governmental. Also, there exists a large number of information sources regarding cryptocurrencies, which consequently leads to the formation of diverse opinions or clusters concerning investments about cryptos.

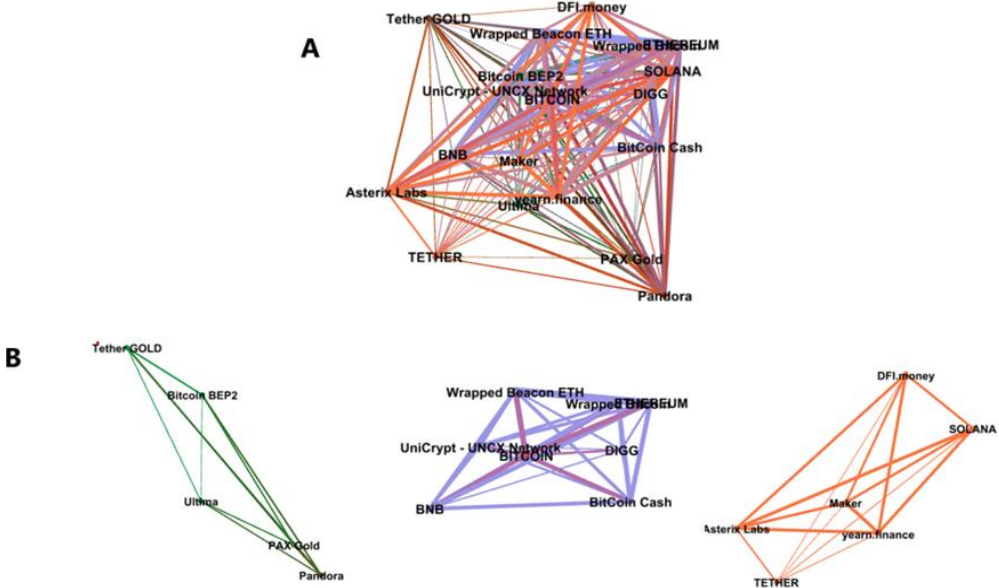

*Figure 2*: *(A) Network of all cryptocurrencies analyzed based on data to 2024 (B) Three communities separately.*

Additionally, it is observed that the nodes are uniform in size (figure 3), indicating that all cryptocurrencies are interconnected with one another. This same pattern is evident in the three communities, where the nodes also have the same size.

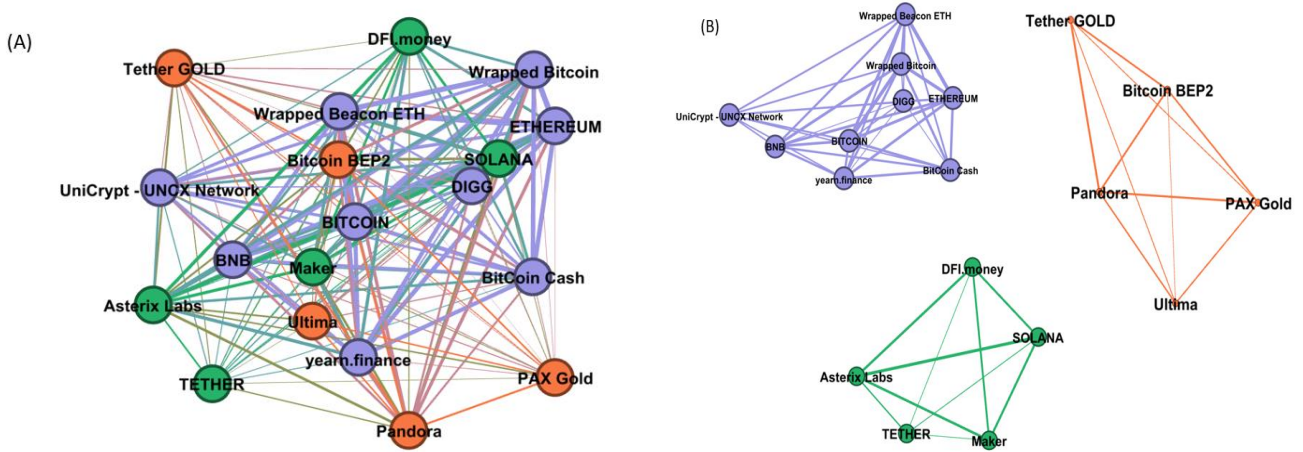

***Figure 3***: *(A) Network with all data to 2024 according the weights (B) Weights of three communities according to the weights from all data.*

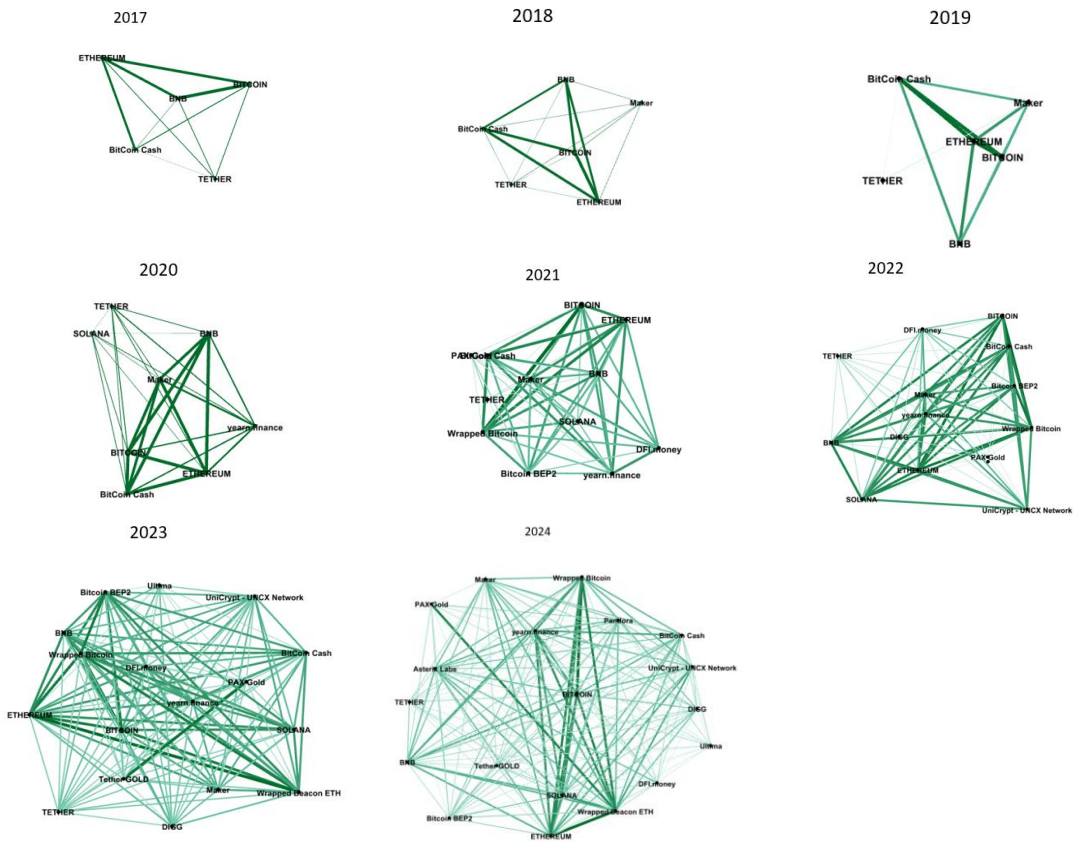

***Figure 4***: *All networks from 2017 to 2024 according to their degree and the number of nodes*

Also, this study examines the evolution of cryptocurrencies from 2017 to 2024. It is observed that there was a notable increase in the number of nodes during the COVID-19 pandemic, a trend that persisted through to 2024. It is observed that as new cryptocurrencies imported by the time to the network, there are changes to the existing correlations. There is also a rise in closeness centrality, betweenness centrality, and eigenvector centrality. Conversely, the metrics of Network Diameter, Graph Density, and Average Path Length remain stable.

*3.2. Data Collection and Discriminant Analysis*

The alternative strategy involved gathering not only the daily closing prices of cryptocurrencies but also additional data points such as the opening price, the highest price, the lowest price, and the trading volume. This approach could uncover any patterns or groupings present within the nineteen cryptocurrencies. It is certainly to determine whether there are any groups or clusters that have emerged during the COVID-19 pandemic or during the war in Ukraine. Due to our multivariate data, it is preferable to use a statistical tool like discriminant analysis, which can combine all these features into matrices pairwise and effectively represent the clusters. So, we can either confirm or deny the differences among cryptocurrencies. As time progresses, it becomes evident that cryptocurrencies exhibit distinct characteristics. Notably, in 2024, Solana stands out as different from other cryptocurrencies in terms of opening price, the highest price, the lowest price, and the trading volume. It is possible that Solana appeals to a different demographic of investors, and exploring the characteristics of this group could contain valuable insights.

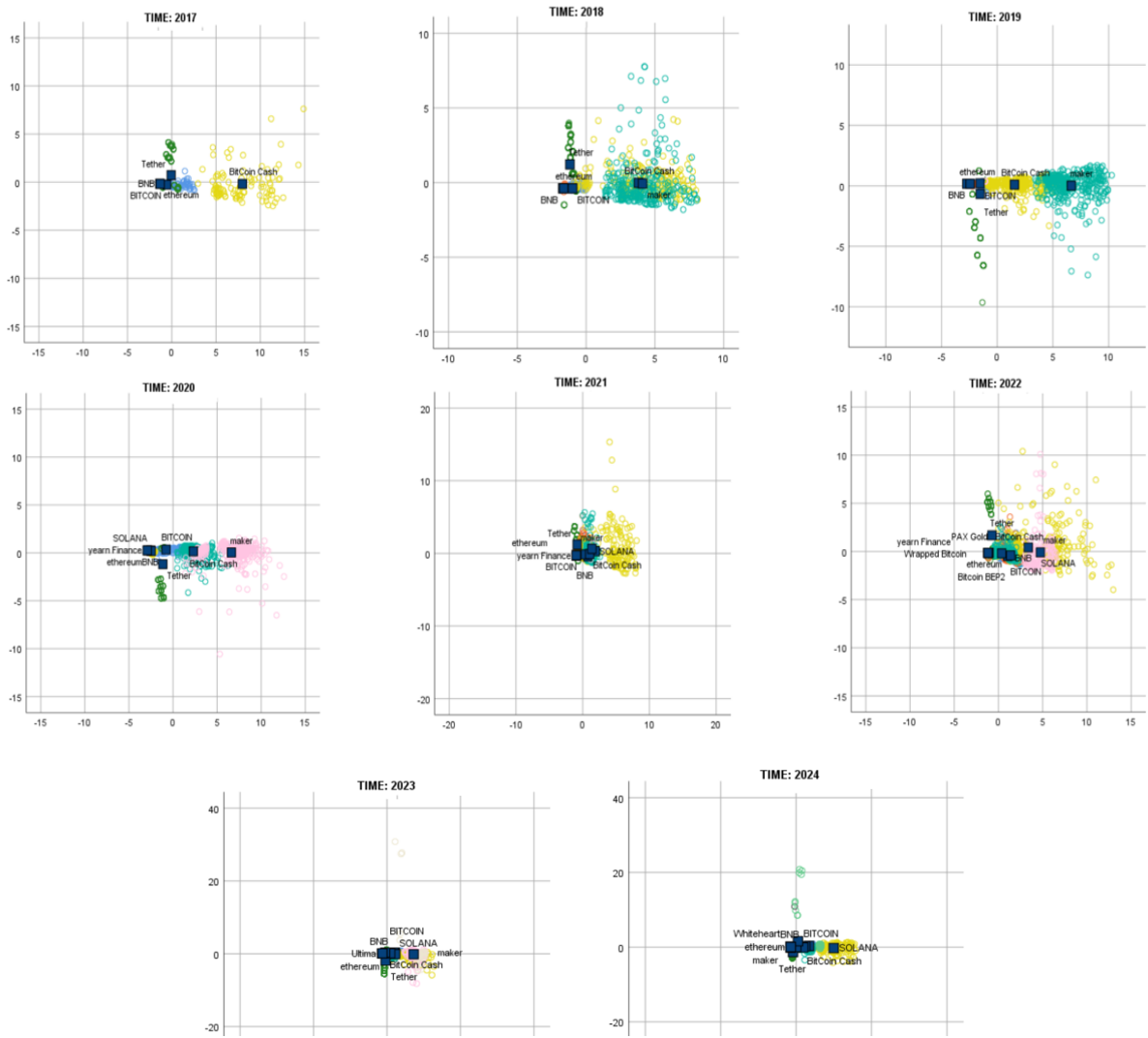

*Figure 5* *Discriminant analysis of cryptos according to their features per time from 2017 to 2024*

## 4. Conclusions

In the present work we studied the cryptocurrency ecosystem using Complex Networks and Discriminant Analysis. The main results are the follows.

- The identification of **three distinct communities** supports the idea that the system is composed of different interacting subsystems with unique characteristics and dynamics. Also, provide evidence of the existence of three different clusters, each accompanied by their respective "fans" of investors.
- The Covid-19 crisis and Ukraine war are widely recognized for their significant impact on the financial markets. Nevertheless, this study reveals an increase in the number of communities, the average degree, the number of nodes, and the number of edges, suggesting that cryptocurrencies are not affected by the crisis. This is proved by the fact that during the period from 2017 to 2024 there has been a notable increase in their market capitalization, indicating a growing trust among investors.[17]
- For the same period, it has been found that the results from complex networks and discriminant analysis are consistent concerning the existence of communities among cryptocurrencies.
- The observed increase of betweenness centrality indicates that network is becoming more efficient in the dissemination of information, since the above increase is related with the decrease of the distances among various communities [18].
- The increase in eigenvector centrality indicates an expansion of the network [19], reflecting a growing number of users or entities linking to key nodes or influential participants within the ecosystem over the time.
- The increase of closeness centrality during the period from 2017 to 2024 reveals a similar behavior with the confidence of investors, leading them to invest in a wide array of digital currencies. [20].
- Future studies may investigate the relationships among these clusters, incorporating additional elements of international financial system such as gold and oil prices.